%% file: main.tex
\documentclass[12pt]{article}

\usepackage{url,hyperref,lineno,microtype,subcaption}
\usepackage[onehalfspacing]{setspace}
\usepackage{pdfpages}

\usepackage{amsmath}
\usepackage{MnSymbol}%
\usepackage{wasysym}%

\usepackage{graphicx,scalerel}
\newcommand\sbullet[1][.5]{\mathbin{\ThisStyle{\vcenter{\hbox{%
  \scalebox{#1}{$\SavedStyle\bullet$}}}}}%
}
\usepackage{xcolor}
\usepackage{soul} 
\usepackage{microtype}

\definecolor{ntrivial}{RGB}{129,60,45}
\definecolor{darkorange}{RGB}{221,110,78}
\definecolor{darkred}{RGB}{153,51,51}
\definecolor{darkgray}{RGB}{77,75,75}
\definecolor{midblue}{RGB}{1,128,255}
\definecolor{darkblue}{RGB}{0,51,102}
\definecolor{lime}{RGB}{0,233 ,0}
\definecolor{greenblue}{RGB}{0,142,100}
\definecolor{darkgreen}{RGB}{0,102,51}
\definecolor{midgreen}{RGB}{76,153,0}
\definecolor{dead}{RGB}{30,144,255}
\definecolor{fix}{RGB}{50,205,50}
\definecolor{cyclic}{RGB}{255,215,0}
\definecolor{chaos}{RGB}{255,69,0}
\definecolor{ns}{RGB}{139,69,19}
\definecolor{classicgreen}{RGB}{0, 128, 0}
\definecolor{classicred}{RGB}{255, 0, 0}
\definecolor{classicorange}{RGB}{255, 165, 0}
\definecolor{firebrick}{RGB}{178, 34, 34}
\definecolor{black}{RGB}{0, 0, 0}
\usepackage{booktabs}
\usepackage{booktabs}
\usepackage{colortbl,hhline}

\usepackage{natbib}
\newcommand{\beginsupplement}{%
        \setcounter{table}{0}
        \renewcommand{\thetable}{S\arabic{table}}%
        \setcounter{figure}{0}
        \renewcommand{\thefigure}{S\arabic{figure}}%
     }

\title{The connectivity degree controls the difficulty of RBN reservoir design}

\author
{Emmanuel Calvet$^{1\ast}$, Bertrand Reulet$^{2}$, Jean Rouat$^{1}$\\
\\
\normalsize{$^{1}$\href{https://www.gegi.usherbrooke.ca/necotis/?lang=en}{NECOTIS}, Génie électrique, Université de Sherbrooke, Canada}\\
\normalsize{$^{2}$\href{https://www.usherbrooke.ca/iq/en/}{Institut Quantique}, Département de Physique, Université de Sherbrooke, Canada}\\
\\
\normalsize{$^\ast$E-mail:  emmanuel.calvet@usherbrooke.ca}
}

\date{}

\begin{document} 
\setlength{\belowdisplayskip}{0.1pt} \setlength{\belowdisplayshortskip}{0.1pt}
\setlength{\abovedisplayskip}{0pt} \setlength{\abovedisplayshortskip}{0pt}
\appto{\bibsetup}{\raggedright}


\baselineskip24pt


\maketitle

\setstretch{1.20}

\begin{abstract}
Reservoir Computing (RC) is a paradigm in artificial intelligence where a recurrent neural network (RNN) is used to process temporal data, leveraging the inherent dynamical properties of the reservoir to perform complex computations. In the realm of RC, the excitatory-inhibitory balance $b$ has been shown to be pivotal for driving the dynamics and performance of Echo State Networks (ESN) and, more recently, Random Boolean Network (RBN). However, the relationship between $b$ and other parameters of the network is still poorly understood. This article explores how the interplay of the balance $b$, the connectivity degree $K$ (i.e., the number of synapses per neuron) and the size of the network (i.e., the number of neurons $N$) influences the dynamics and performance (memory and prediction) of an RBN reservoir. Our findings reveal that $K$ and $b$ are strongly tied in optimal reservoirs. Reservoirs with high $K$ have two optimal balances, one for globally inhibitory networks ($b<0$), and the other one for excitatory networks ($b>0$). Both show asymmetric performances about a zero balance. In contrast, for moderate $K$, the optimal value being $K=4$, best reservoirs are obtained when excitation and inhibition almost, but not exactly, balance each other. For almost all $K$, the influence of the size is such that increasing $N$ leads to better performance, even with very large values of $N$. Our investigation provides clear directions to generate optimal reservoirs or reservoirs with constraints on size or connectivity.
\end{abstract}

\input{introduction}

\input{methodology}
\input{results}

\input{conclusion}

\section*{Ackowledgement}

The authors are grateful to Lucas Herranz for carefully reviewing the manuscript. They also want to thank their colleagues at NECOTIS for their helpful feedback and productive discussions during the research process.

\section*{Conflict of Interest Statement}
The researchers affirm that their study was carried out without any affiliations or transactions that could give rise to a possible conflict of interest.

\section*{Author Contributions}

This article is a collaboration between EC, BR, and JR. EC conducted the research, developed the model, collected and analyzed data, and wrote the manuscript. BR and JR provided guidance, perspective, and contributed to the manuscript. All authors approved the final version.

\section*{Funding}

This work was supported by the CRSNG/NSERC, the Canada Research Chair Program, NSERC, and CFREF.

\section*{Code and data availability}

The study used original code and data created by the author. The repository \href{https://zenodo.org/doi/10.5281/zenodo.10247106}{DOI:10.5281/zenodo.10247106} contains the code and data used to perform the simulations.

\bibliography{bibliography}

\input{supplementary}

\end{document}

%% file: introduction.tex
\vspace{10pt}
\section{Introduction}\label{Sec:Introduction}

Reservoir computing (RC) is a promising approach that could drastically reduce the cost of learning as the input gets projected into a higher dimensional space, \textit{the reservoir}, read out by a single output layer. As such, when the reservoir is adequately designed, a simple linear fitting can be used to train the weights of the readout layer \citep{Maass2002b}, alleviating the computational burden of other traditional machine learning methods. The Echo State Network (ESN) developed by \citep{Jaeger2005} comprises reservoirs with continuous activation functions, while Liquid State Machine (LSM) \citep{Maass2002b} typically includes discontinuous activation functions, among which we find the Random Boolean Network (RBN) \citep{Glass1998}.

The connectivity degree has been extensively studied in RBN reservoirs, and in contrast to ESN, it displays desirable dynamics for very sparse matrices \citep{Luque2000, Bertschinger2004c, Busing2010, Snyder2012, Echlin2018}, while ESN was shown to be less sensitive to this parameter \citep{Hajnal2006, Busing2010, Krauss2019a, Metzner2022}. On the other hand, it is well known that increasing the number of neurons improves performance \citep{Bertschinger2004c, Snyder2012, Cherupally2018, Cramer2020, Steiner2022}. However, most literature on RBN compared reservoirs with rather small sizes around $1000$ neurons \citep{Bertschinger2004c, Natschlager2005, Busing2010, Snyder2013a, Burkow2016}, while studies on the ESN compared reservoirs from $500$, up to $20000$ neurons \citep{Triefenbach2010}.

In this article, we want to study the effect of these topology parameters ($N$ and $K$), with another control parameter, which is the excitatory-inhibitory balance $b$, controlling the proportion of positive and negative synaptic weights \citep{Krauss2019a, Metzner2022, Calvet2023}. The balance is equal to $b=(S_+ - S_-)/S$, with $S=KN$ the total number of synapses and $S_\pm$ the number of positive and negative synapses. For a positive balance, the network has a majority of excitatory synapses and reverse, and when it is zero, the network has a perfect balance between the two, $S_+=S_-$. Previous work on the ESN \citep{Krauss2019, Krauss2019a, Metzner2022} has studied the influence of density $d=K/N$ and balance on the dynamics of reservoirs, showing that $b$ was a key parameter controlling phase transitions. In particular, the edge of chaos, a dynamical phase transition between order and chaos, is believed to be fundamental in reservoir design, for a reservoir to acquire computational capabilities (for a comprehensive introduction, see \citep{Legenstein2006, Nowshin2020}). Nevertheless, Metzner \textit{et al.} suggested a more complex picture than previously thought, exposing two critical points, each for a positive and negative balance, while for higher densities, an asymmetry could arise in the reservoir responses to inputs, and as a result, only the edge of chaos occurring for positive $b$ was optimal for information propagation inside the reservoir \citep{Metzner2022}. 

In line with Krauss and Metzner, recent work on RBN reservoirs demonstrated that the excitatory-inhibitory balance $b$ was also key in driving dynamics and performance \citep{Calvet2023}. In particular, it was shown that the weight statistics, typically used in RBN literature \citep{Bertschinger2004c, Natschlager2005, Busing2010} are related to the balance. More striking, the RBN reservoirs also displayed an asymmetry around $b=0$. The two signs of the balance produced distinct relations to performance in tasks and a reduced reservoir-to-reservoir variability for a majority of inhibition. However, this occurred for a network with extremely low density as $d=K/N=16/10000=0.0016$, in contrast with studies on ESN.

As far as the authors are aware, the influence of the excitatory-inhibitory balance for different connectivity has yet to be studied, except for the single value of $K=16$ previously mentioned \mbox{\citep{Calvet2023}}. This article aims to explore the combined effect of connectivity ($K$, $N$) and the balance on the dynamics and performance of the RBN. The article is organized as follows: in the first section (\mbox{Sec.~\ref{sec:K}}) the effect of $K$ and $b$ is studied, both on the dynamics of free-evolving reservoirs (\mbox{Sec.~\ref{sec:dynamic_K_and_b}}), and their performance in a memory and prediction task (\mbox{Sec.~\ref{sec:perf_K_and_b}}), showing that the asymmetry in fact vanishes for very small $K$. In the second section (\mbox{Sec.\ref{sec:N}}) we perform a similar analysis (dynamics in \mbox{Sec.~\ref{sec:dynamic_K_and_b}}, and performance in \mbox{Sec.~\ref{sec:performance_K_and_N}}), but this time, we vary both $K$ and $N$ conjointly, and explore the relationship with $b$. This reveals a complex interplay between parameters and suggests that $K$ is, in fact, governing it. Finally, in \mbox{Sec.~\ref{Sec:Conclusion}}, we discuss our results and their implication for RBN reservoir design, revealing that in contrast with $ESN$, the careful selection of $K$ leads to a significant simplification of the fine-tuning of the other topology parameters in the tested tasks.

%% file: methodology.tex
\vspace{10pt}
\section{Methodology}\label{met:main}

\subsection{The model} \label{met:model}

\begin{figure}
    \centering
    \includegraphics[scale=0.3]{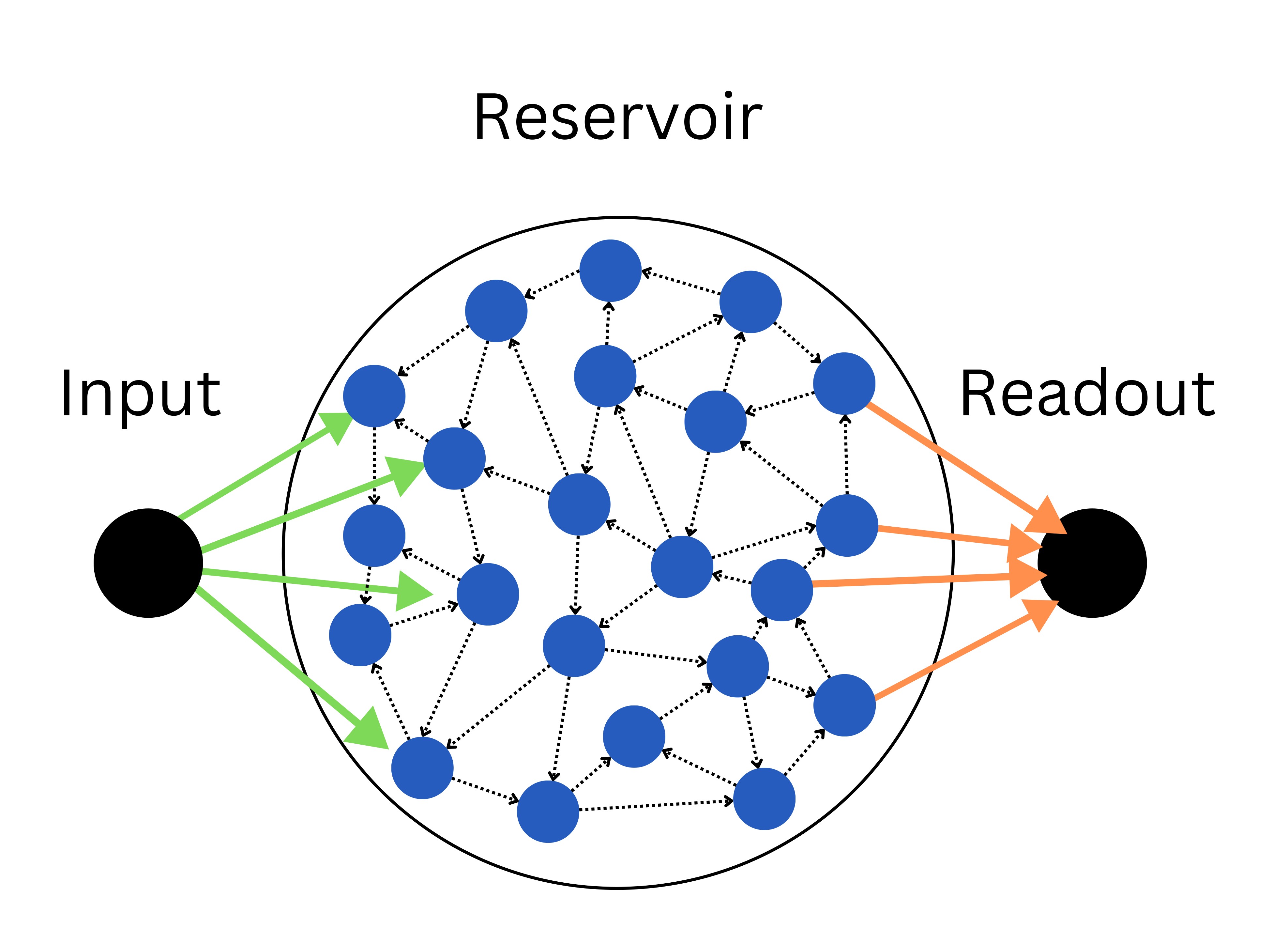}
    \caption{The model consists of an input node (left), connected by input weights (green arrows), to the reservoir (center), itself connecting via output weights (orange arrows) to the output node (right). As illustrated by the dotted black arrows, the reservoir is recurrently connected, forming a random graph. The illustrated graph has $K=2$ and $N=22$. Note that in practice, half of the neurons (blue circles) connect to the input, and the other half to the readout.}
\label{fig:schema}
\end{figure}

Our model is an ensemble of three parts (\mbox{Fig.~\ref{fig:schema}}), the input node $u(t)$, which is projecting to half of the neurons of the recurrently connected reservoir $\vec{x}$, among which the other half is projecting to the output node $y(t)$, this way, the output node never directly sees the input, and information must propagate inside the reservoir for the readout to accomplish the task at hand:

\begin{flalign}
        u_i(t) = w^{in}_i u(t) \\
        y(t) = f(W^{out} \vec{x} + c) 
\end{flalign}

\noindent With $u_i(t)$ the input of the neuron $i$, the input weights $w_i^{in}$ form a vector, projecting to half of the reservoir, while the other half of the weights are zeros, and reserve for the output weight matrice $W^{out}$. This way, a neuron in the reservoir is never connected to both the input and output. The activation function $f$ of the output node is the sigmoid, with a bias $c$. Each component $x_i(t)$ of $\vec{x}(t)$ corresponds to the state of the neuron $i$ inside the reservoir. It is given by:

\begin{equation}
        x_i(t)=  \theta \left(u_i(t) + \sum_{j=1}^N w_{ij}x_j(t-1) \right)  
\end{equation}

\noindent Where each neuron is connected to $K$ other neurons, and $w_{ij}$ is the synaptic weight connecting neuron $j$ to neuron $i$, drawn in a normal distribution $\mathcal{N}(\mu, \sigma)$, with parameters $\mu$ (mean) and $\sigma$ (standard deviation). The activation function $\theta$ is a Heaviside, thus $x_i$ is binary. $t\in \mathbf{N}$ and    corresponds to a time step. Remark that if the input is zero, the state of a given neuron only depends on the states of its neighbours at the previous time step. Such neurons are thus said to be "memoryless", and for such a system, to sustain memory, information needs to cascade via the propagation of spikes inside the reservoir. The attractive feature of the reservoir framework is that only the output weight and bias are trainable parameters, as all other parameters are usually kept fixed, including the reservoir weights. 

We use a mean square error (MSE) loss function for the training process. For training the readout weights, we opted for the ADAM optimizer \citep{Kingma2015}, providing superior results in our testings, superseding the commonly utilized Ridge regression \citep{Burkow2016} in most literature. The execution is facilitated through the PyTorch library, with parameters set at $\alpha=0.001$ and $4000$ epochs (Supplementary Materials \ref{suppm:training} for additional information).

\subsection{The control parameters} \label{met:control_param}

\begin{figure}
    \centering
    \includegraphics[scale=0.5]{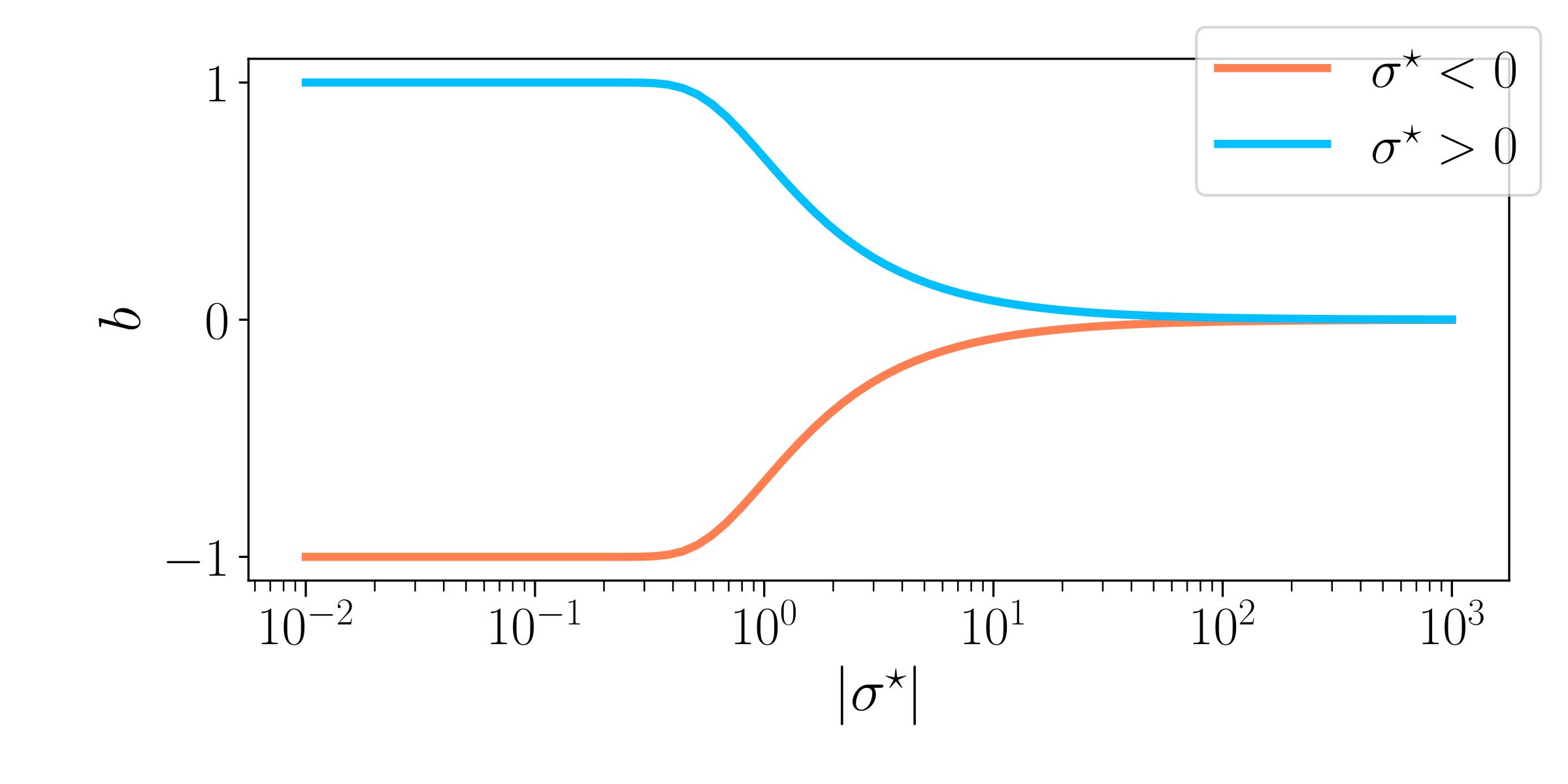}
    \caption{The excitation-inhibition balance $b$ as a function of the synaptic weight parameter $\sigma^\star$. For $\sigma^\star<0$ (\textcolor{cyan}{\rule[0.5ex]{1em}{0.6pt}}) and $\sigma^\star>0$ (\textcolor{darkorange}{\rule[0.5ex]{1em}{0.6pt}}). $\sigma^\star$ is the coefficient of variation ($\sigma/\mu$) of the weight distribution, which is why $b$ is of the sign of $\sigma^\star$. For low values of $|\sigma^\star|$, only $\mu$ controls the synaptic balance, meaning that for a positive mean, the weights are all excitatory, and reverse. On the other hand, when $|\sigma^\star| \rightarrow \infty$, the mean becomes irrelevant, and $b$ is at a perfect balance between excitation and inhibition.}
\label{fig:balance}
\end{figure}

The three control parameters used in this study are $\sigma^\star$, $K$, and $N$. Among these, $\sigma^\star$ represents the coefficient of variation of the weight distribution within the reservoir, defined as $\sigma^\star=\sigma/\mu$. This parameter is linked to $b$, the excitatory/inhibitory balance, as $b=\mathrm{Erf}[1/(\sqrt2\sigma^\star)]$ \citep{Calvet2023}. The balance is also equal to $b=(S_+ - S_-)/S$, with $S$ the total number of synapses, and $S_\pm$ the number of positive and negative synapses, respectively. We display in Fig.~\ref{fig:balance} the relationship between the two, noting that when $\sigma^\star$ is positive, we have a majority of excitatory synapses $b>0$, and when $\sigma^\star$ is negative, we have a majority of inhibitory synapses $b<0$. In all experiments, we play with values of $\sigma^\star$ that allow our reservoirs to span the full range of $b$, corresponding to $\sigma^\star \in [10^{-2}, 10^3]$.

Since recent work showed that the dynamics and performance of reservoirs were asymmetric about $b=0$ \citep{Metzner2022, Calvet2023}, we study the influence of two other control parameters with respect to the sign of $b$. These parameters are captured by the density $d=K/N$, following the work of \citep{Hajnal2006, Krauss2019a, Metzner2022} on ESN. However, we show in supplementary material \ref{suppm:density} that the density $d$ is not a control parameter for the RBN, since, at a fixed density, reservoirs can possess very different dynamics as $K$ and $N$ are concurrently varied. As such, we consider them as independent control parameters in this article. Following work in RBN \citep{Busing2010, Calvet2023}, the connectivity degree is chosen between $1$ and $16$. In addition, to compare the more recent results ($N=10000$) \citep{Calvet2023} with older literature  ($N\le1000$) \citep{Bertschinger2004c, Natschlager2005, Busing2010, Snyder2013a, Burkow2016}, we study three values of $N=\{100, 1000, 10000\}$.

\subsection{The experiments} \label{met:experiments}

We perform two types of tasks: the first to probe the intrinsic dynamics of reservoirs, while they are freely evolving, and the second to test the ability to process inputs while performing memory and prediction tasks.

\subsubsection{Free-running}

Each reservoir is freely running without input for a duration of $D=2000$ time steps, with a random initial state with 20\% of neurons to one. During a run, the activity signal $A(t)$ (the average of states $x_i$ at a given time step $t$) is recorded. Afterward, we compute the BiEntropy ($H_b$) \citep{Croll2014} of the binarized activity signals, which quantify the degree of order and disorder of a bit string, $H_b=0$ for completely periodic, and $H_b=1$ for totally irregular. For each triplet ($N$, $K$, $\sigma^\star$), we randomly generate $100$ reservoirs, and we then compute the average and variance over reservoirs having the same control parameters. 

Next, we classify the steady-state activity $A(t)$, for $t>1000$ time steps, into four distinct attractor categories. For each triplet ($N$, $K$, $\sigma^\star$), we then compute the histograms over the $100$ reservoirs. The attractors are defined according to \citep{Calvet2023} : 

\begin{itemize}
    \item \textbf{Extinguished}: The activity has died out, and the steady activity is zero at all time steps. 
    \item \textbf{Fixed attractor}: The steady activity is non-zero, but its derivative is zero at all time steps. 
    \item \textbf{Cyclic}: The steady activity repeats, with a period larger than one time-step. 
    \item \textbf{Irregular}: If none of the above categories apply, the signal is irregular. Note that our model is deterministic and discrete, as such, all attractors are in theory, cyclic; however, since the duration $D=2000$ is extremely small compared to the maximal period of $2^N$, in practice, we find a statistically significant proportion of attractors in that category. 
\end{itemize}

\subsubsection{Performance in tasks}

To test the computational capabilities of our reservoirs, we perform two distinct tasks. The first one consists of memorizing white-noise input received $|\delta|$ time steps in the past. We test our reservoirs with various difficulties for $\delta=\{-18, -14, -10, -6, -2$\}. The higher in absolute value, the more difficult the task, since it demands the reservoir of memoryless neurons to integrate and reverberate input information through spikes cascade for longer time scales \citep{Metzner2022, Calvet2023}. The second task consists of predicting Mackey-Glass time series, $\delta=10$ time steps in the future. Mackey-Glass is a common benchmark in reservoir computing \citep{Hajnal2006, Bianchi2016, Zhu2021}. We use $\tau$, the time constant parameter of Mackey-Glass, which controls the signal dynamics, ranging from $\tau=5$ (periodic), $\tau=15$, to $\tau=28$ (chaotic). To evaluate the performance of our reservoir, we compute the correlation coefficient $Corr(y, T)$ between the target $T$, and the output $y$ (implementation details are identical to this study \citep{Calvet2023}).

%% file: results.tex
\vspace{10pt}
\section{Results}\label{Sec:Results}

 \subsection{The connectivity degree controls the optimal balance} \label{sec:K}

In this section, we fix the size of the reservoir to its largest value $N=10000$. We study the effect of $K$ and $b$ on the dynamics of free-running reservoirs (Sec.~\ref{sec:dynamic_K_and_b}). Then, we study the performance in two demanding tasks (Sec.~\ref{sec:perf_K_and_b}). We show that the asymmetry about $b=0$ is strongly $K$ dependent and vanishes for low $K$, while the optimal balance $b_{opt}$ is entirely controlled by $K$.   

Additionally, we exhibit the shift of control parameters from the more natural weight distribution statistics ($\sigma^\star$) \citep{Calvet2023} to the excitatory-inhibitory balance ($b$). To do so, we begin by exhibiting the dynamics over $\sigma^\star$, to then display the attractor statistics over the excitatory balance $b$, revealing insights into the reservoir design. 

\subsubsection{Impact of the connectivity degree and balance on dynamics} \label{sec:dynamic_K_and_b}

\begin{figure}
    \centering
    \includegraphics[width=1.1\textwidth]{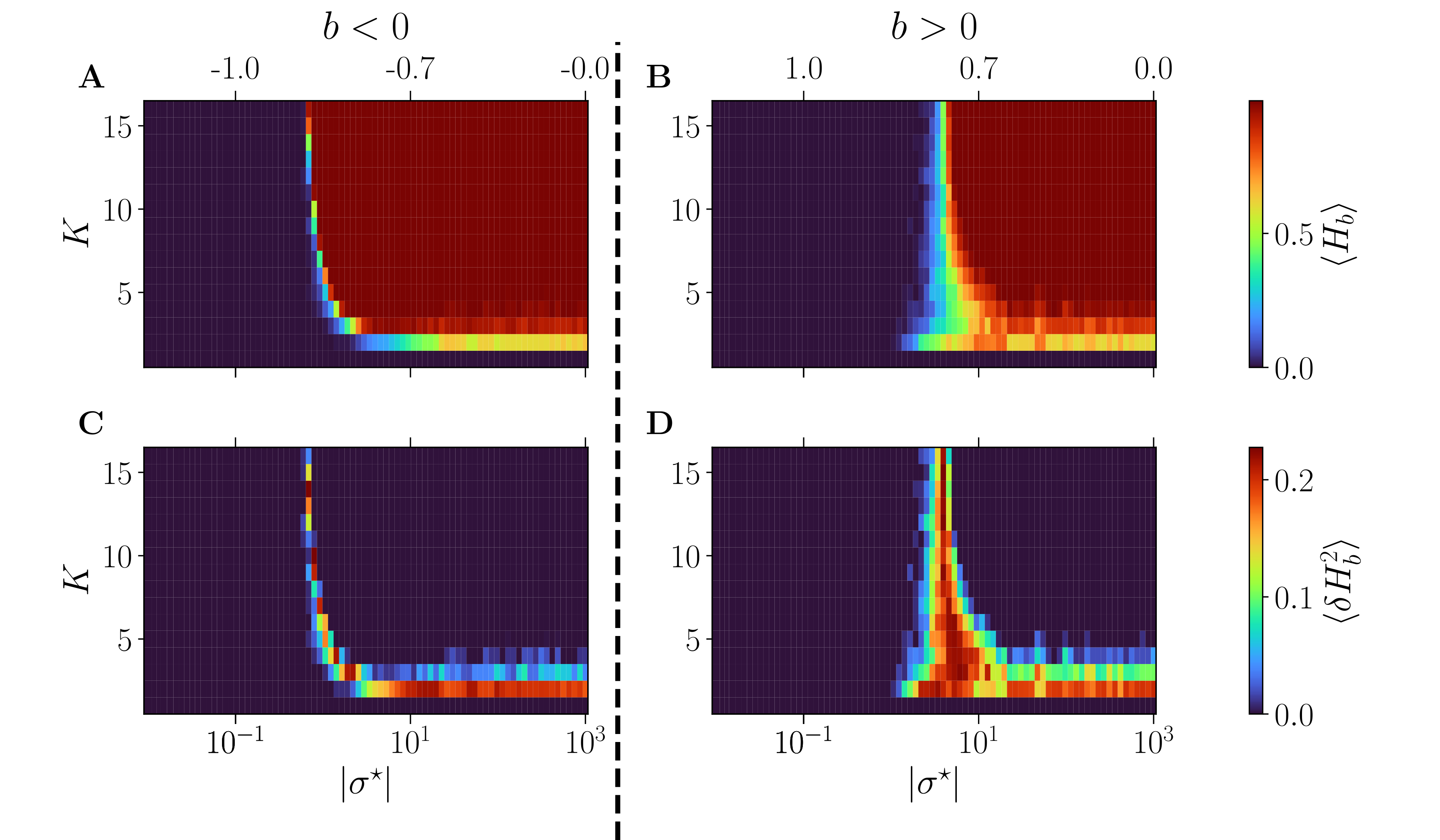}
    \caption{Dynamics of free evolving reservoirs as controlled by the connectivity degree $K$ (y-axis) and $|\sigma^\star|$ (x-axis). The upper x-axis displays the corresponding $b$ values, for $b<0$ (\textbf{A} and \textbf{C}), and $b>0$ (\textbf{B} and \textbf{D}). The BiEntropy is computed on the steady activities of 100 reservoirs per couple ($K$, $\sigma^\star$). (\textbf{A} and \textbf{B}): The upper row displays the average BiEntropy of the steady state activities (upper left colormap). (\textbf{C} and \textbf{D}): The lower row shows the variance of BiEntropy over reservoirs (bottom left colormap).}
    \label{fig:panel1}
\end{figure}

In Fig.~\ref{fig:panel1}, we display the average over reservoirs of the BiEntropy of the steady activities for reservoirs as a function of $|\sigma^\star|$ (lower x-axis), both with a negative (left) or positive (right) balance $b$ (the upper x-axis displays the corresponding $b$ values). In Fig.~\ref{fig:panel1}.\textbf{A} and \textbf{B}, blue regions represent an ordered phase with low BiEntropy, and red regions represent a disordered phase with a BiEntropy close to one. The regions are separated by a phase transition where the BiEntropy is intermediate, also captured by the variance of the BiEntropy (Fig.~\ref{fig:panel1}.\textbf{C} and \textbf{D}). The scenario is similar for both signs of $b$ but differs in the details. The transition (abrupt for $b<0$, wider for $b>0$) occurs at a value of $\sigma^\star$ that depends on $K$ (strongly for $b<0$, weakly for $b>0$). The transition widens when $K$ decreases (strongly for $b>0$). At high $K$, i.e. when each neuron is connected with many, there seems to be an asymptotic value for $\sigma^\star$ (or $b$, indicated on the upper part of the plots), which is different for $b>0$ and $b<0$ \citep{Calvet2023}. For $K=2$, the disordered phase never reaches a BiEntropy of $1$, and for $K=1$ the reservoir is always in its ordered phase \citep{Bertschinger2004c}.

\begin{figure}
    \centering
    \includegraphics[width=0.9\textwidth]{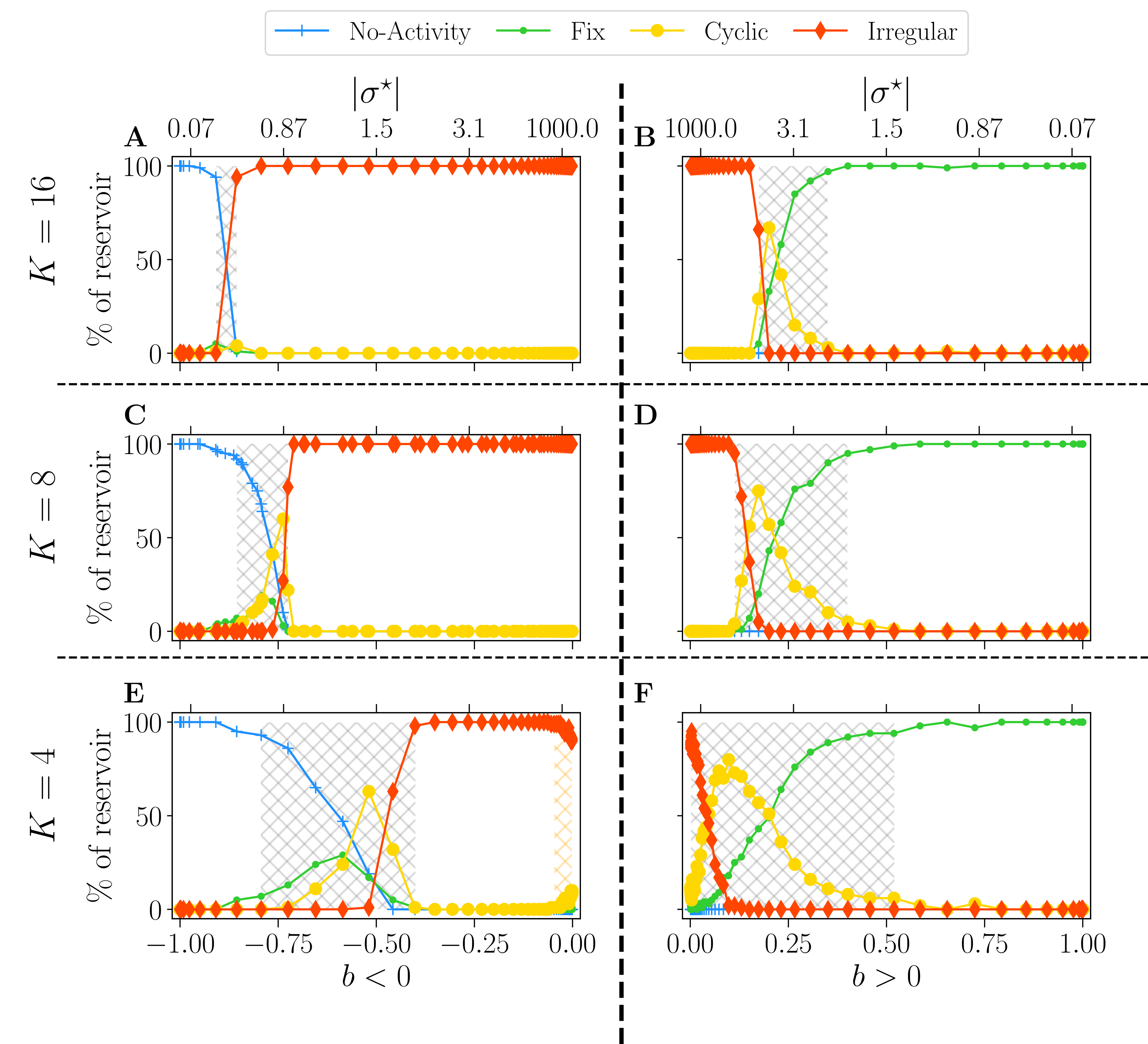}
    \caption{Attractor statistics of free-evolving RBN reservoirs, controlled by $K$ (rows), and the balance $b$ (x-axis). The upper x-axis represents the corresponding $|\sigma^\star|$ values, for $b<0$ (\textbf{A}, \textbf{C}, \textbf{E}), and $b>0$ (\textbf{B}, \textbf{D}, \textbf{F}). All reservoirs are of size $N=10000$. Each steady activity signal is classified into one of the four categories of attractors: no-activity (\textcolor{dead}{$+$}), fix (\textcolor{fix}{$\sbullet$}), cyclic (\textcolor{cyclic}{$\bullet$}), irregular (\textcolor{chaos}{$\blacklozenge$}). The statistics of attractors are computed over 100 reservoirs run once (y-axis). Results are shown for $K=16$ (\textbf{A} and \textbf{B}), $K=8$ (\textbf{C} and \textbf{D}), and $K=4$ (\textbf{E} and \textbf{F}). The light-gray hatched areas represent the critical regions \citep{Calvet2023}, defined as the region of non-zero BiEntropy variance; the threshold is chosen to $0.0001$. In \textbf{E}, the orange hatched area represents a region of re-entrance of criticality with non-zero BiEntropy variance, distinct from the critical region. All hatched areas are computed from the data shown in Fig.~\ref{fig:panel1}.\textbf{C} and \textbf{D}.}
    \label{fig:panel2}
\end{figure}

In Fig.~\ref{fig:panel2}, we plot the statistics of attractors for reservoirs with $K=16$ (upper panel), $K=8$ (middle), and $K=4$ (lower), as a function of the balance $b$. This time $|\sigma^\star|$ is reported in the upper x-axis. The left column shows the results for $b<0$ and the right column for $b>0$. The phase transition is characterized by going from attractors with essentially no ($b<0$) or fixed ($b>0$) activity in the ordered phase, to attractors being all irregulars in the disordered phase, with cyclic attractors showing up at the transition. In all plots, we report the non-zero BiEntropy variance (highlighted by light-grey hatching) to indicate the critical region \citep{Calvet2023}. This transition region is clearly defined for $K=16$, widens for $K=8$ and becomes very different for $K=4$. When $b<0$, there is a transition region around $b\sim-0.7$ (gray hashed region) and a re-entrance of the critical region (orange hatching in Fig.~\ref{fig:panel2}.\textbf{E}). Indeed, for $b$ between -0.7 and -0.08 ($\sigma^\star$ between $-10$ and $-2$) all attractors are irregular, and cyclic ones reappear for a balance closer to zero. For $K=4$ and $b>0$ the phase transition is never complete, there is no fully disordered phase. Lastly, near $b=0$, the attractor statistics are very close from one sign to the other. For example, with $K=16$ and $K=8$ we observe an horizontal line for chaotic attractors, while for $K=4$, the statistics of cyclic and irregular attractors closely match on both sides, a fact that is even more visible in the results of Sec.~ \ref{sec:dynamic_K_and_N} when varying $N$.

Regarding the control parameter shift from $\sigma^\star$ to $b$, the phase transition appears inflated in $b$, as indicated by the dot positions, particularly for $b<0$. These positions are generated on an evenly spaced logarithmic scale in $\sigma^\star$. The irregular regime is notably compressed, demonstrated by the re-entrant critical region (refer to Fig.~\ref{fig:panel2}.\textbf{E}), spanning from $2.10^1$ to $10^3$. This observation suggests that the dynamics remain relatively consistent despite significant variations in the weight distribution parameter. In line with \citep{Metzner2022, Calvet2023}, we make the case that underlying $b$ is what is driving the dynamics of these reservoirs. As such, in the rest of the article, we use $b$ as a reference for all further investigations. 

In conclusion, $K$ has a strong influence on the dynamics of the network. For large values of $K$, a variety of attractors can be found only in a narrow region of $b$ ($\sigma^\star$), which is different for both signs of the balance. In contrast, for lower values of $K$, the co-existence of several attractors is found over a very wide range of $\sigma^\star$ which corresponds to the region where $b$ is small, positive or negative. 

\begin{figure}
    \centering
    \includegraphics[width=1\textwidth]{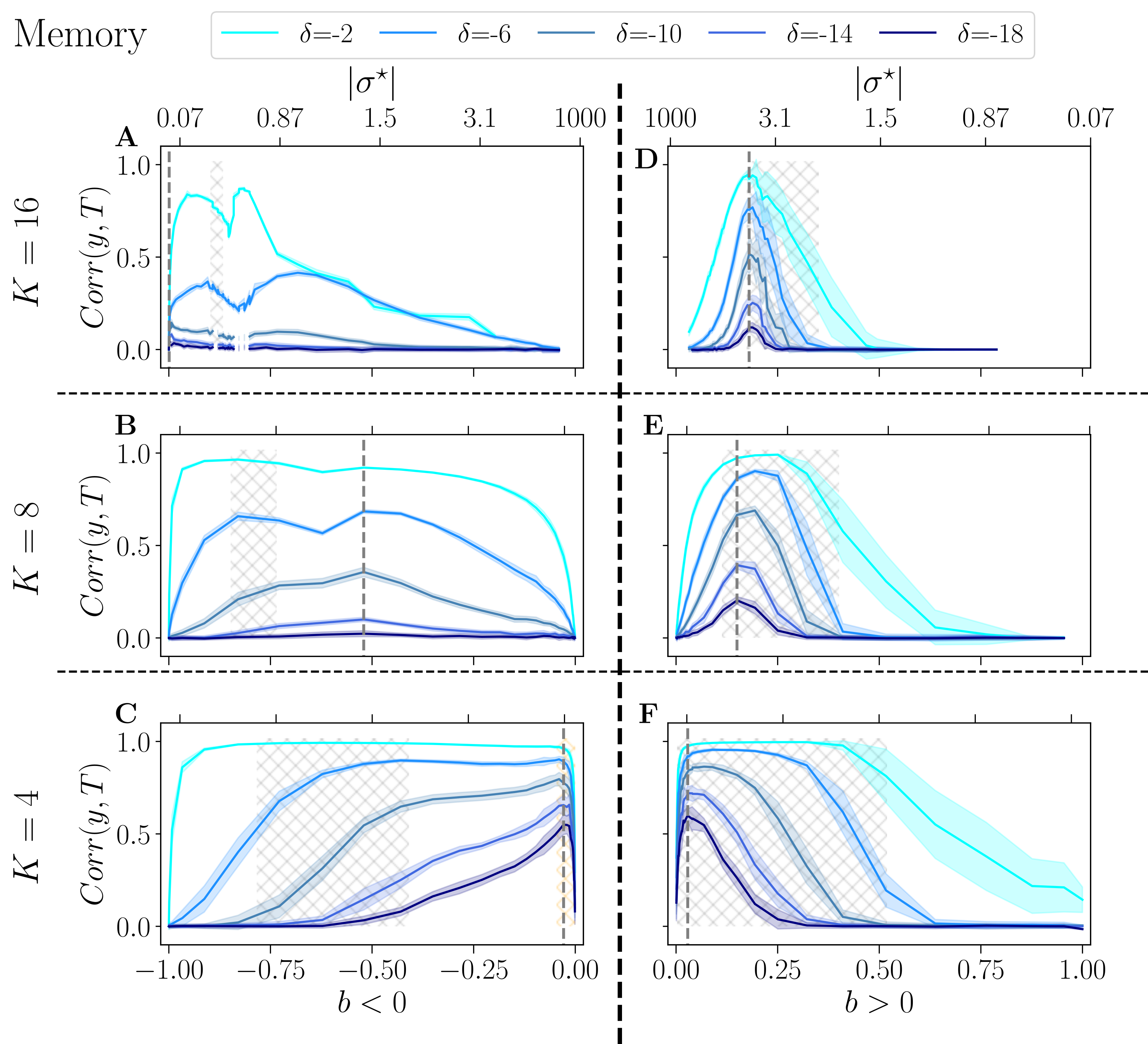}
    \caption{Performance of RBN reservoirs in the memory task of white-noise signals, for various $\delta$, the higher in absolute, the more difficult the task. The correlation between the target and the output (y-axis), is plotted as a function of the control parameter $b$ (x-axis), for a positive balance (\textbf{A}, \textbf{B}, \textbf{C}), and a negative balance (\textbf{D}, \textbf{E}, \textbf{F}). The upper x-axis represents the corresponding $|\sigma^\star|$ values. The solid lines represent the average over $20$ reservoirs, higher values signify better performance, while the shaded area represents one standard deviation. (\textbf{A} and \textbf{E}): the upper row displays $K=16$, the middle row (\textbf{C} and \textbf{F}) $K=8$, and bottom row (\textbf{D} and \textbf{G}) $K=4$. The light-gray hatched areas represent the critical regions of BiEntropy variance above a threshold of $0.0001$, and the dotted gray lines represent the optimal balance $b_{opt}$ in the most difficult task.}
    \label{fig:panel31}
\end{figure}

\subsubsection{Impact of the connectivity degree and balance on performance} \label{sec:perf_K_and_b}

In Fig.~\ref{fig:panel31}, we show the performance of the reservoirs for memory tasks as a function of the control parameter $b$ ($|\sigma^\star|$ upper x-axis). Five difficulties are operated, with $\delta$ varying from $-2$ to $-18$. The left column comprises reservoirs with a negative balance and the right column with a positive one. We show the results for $K=16$ (upper row), $K=8$ (middle), and $K=4$ (bottom). 

\begin{figure}
    \centering
    \includegraphics[width=1\textwidth]{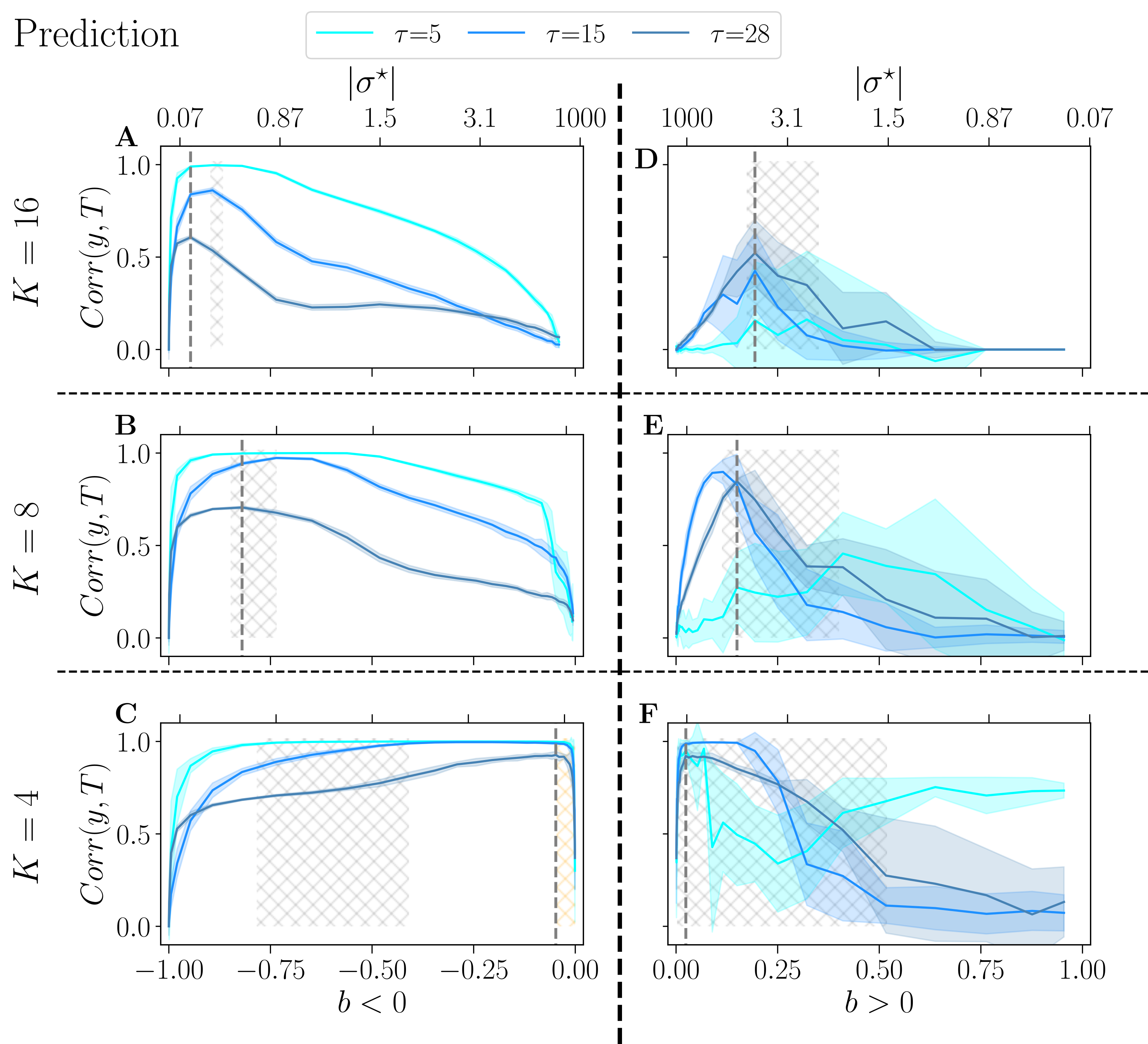}
    \caption{Performance of RBN reservoirs in the prediction task of Mackey-Glass time series, for various $\tau$, the higher, the more complex the signal. The correlation between the target and the output (y-axis), is plotted as a function of the control parameter $b$ (x-axis), for a positive balance (\textbf{A}, \textbf{B}, \textbf{C}), and a negative balance (\textbf{D}, \textbf{E}, \textbf{F}). The upper x-axis represents the corresponding $|\sigma^\star|$ values. The solid lines represent the average over $20$ reservoirs, higher values signify better performance, while the shaded area represents one standard deviation. (\textbf{A} and \textbf{E}) $K=16$, with similar result to \citep{Calvet2023}, (\textbf{C} and \textbf{F}) $K=8$, and (\textbf{D} and \textbf{G}) $K=4$. As in the previous figure, the light-gray hashed areas represent the phase transition region, and the dotted gray lines represent the optimal balance $b_{opt}$ in the most difficult task.}
    \label{fig:panel32}
\end{figure}

For each value of the delay, reservoirs perform better at low $K$, and show good performance over a broader range of $b$. Similar observations have been reported for other tasks \citep{Busing2010}.  The balance for which performance is best $b_{opt}$ (dotted gray line) strongly depends on $K$: this is the most visible for $b<0$ and $\delta=-18$ (the most difficult task), where $b_{opt}$ goes from almost $-1$ for $K=16$, to almost $0$ for $K=4$ (see Tab.~\ref{tab:1}). For other values of $\delta$ the effect is less pronounced but clearly always present. For $b>0$ the same phenomenon appears and $b_{opt}$ shifts from $\sim0.2$ for $K=16$, to $\sim0$ for $K=4$. Thus, the asymmetry between $b>0$ and $b<0$ fades as $K$ decreases. For $K=4$, the optimal balance, whether positive or negative, is almost zero, i.e., it corresponds to an almost perfect balance between excitation and inhibition. However, notes that performance drops abruptly for $b=0$: the unbalance, even very small, is essential.

\begin{figure}
    \centering
    \includegraphics[width=1\textwidth]{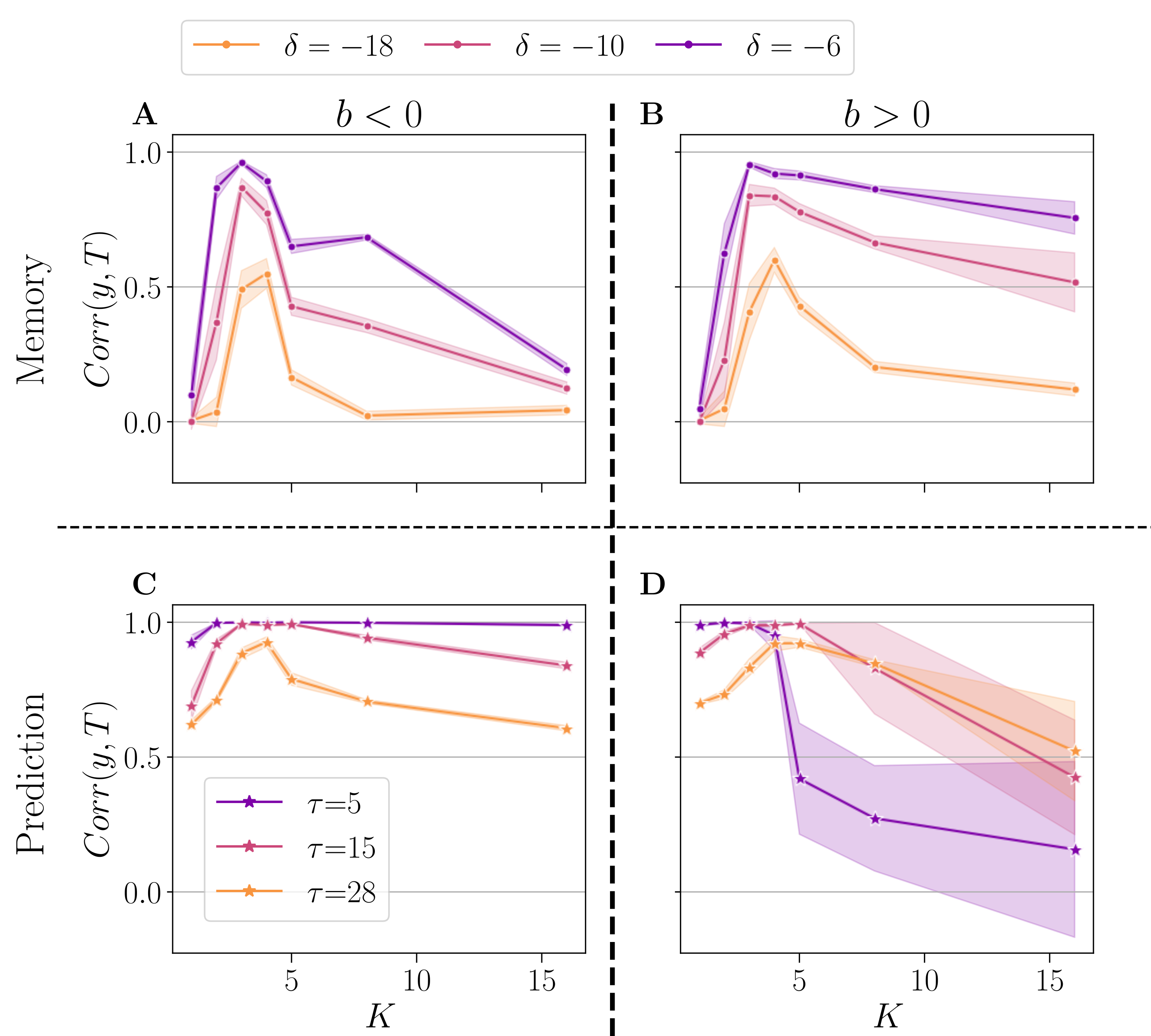}
    \caption{Summary of performance for various connectivity degrees $K$, in the memory (\textbf{A} and \textbf{B}) and the prediction (\textbf{C} and \textbf{D}) tasks. For both $b<0$ (left panel), and $b>0$ (right panel). For each value of $K$, we selected the $b_{opt}$ value giving the highest average performance, in the most difficult task ($\delta=-18$ for memory, and $\tau=28$ for prediction). We plot the performance (higher is better) of reservoirs $Corr(y, T)$ (y-axis), plotted as a function of $K$ (x-axis). The solid line represents the average over $20$ reservoirs (generated with the same $b_{opt}$ and $K$ value), and the shaded area represents one standard deviation. Performance is shown for various $\delta$ in the memory task (\textbf{A} and \textbf{B}), and $\tau$ in the prediction (\textbf{A} and \textbf{B}).}
    \label{fig:panel4}
\end{figure} 

In the prediction task (Fig.~\ref{fig:panel32}), a similar trend is observed: as $K$ decreases, the high-performing region shifts towards $b$ values close to zero. Furthermore, the range of $b$ values within the high-performing region is also broader. Still, for $K=4$, our task may not be sufficiently challenging for the reservoirs, since at $b_{opt}$, the three values of $\tau$ give very close results. When $b<0$, the critical region (gray hashed area) does not align well with the performance peaks, and this discrepancy is even more pronounced for lower $K=4$. The peak of performance is still within the orange-hashed region, indicative of re-entrant criticality. In the case where $b>0$, in line with previous work \citep{Calvet2023}, the variance is exceptionally high, especially for simpler signals $\tau=5$ and $\tau=15$. Surprisingly, for $K=16$ and $K=4$, reservoirs perform better at the complex task than at the simpler task $\tau=5$.

Trying to relate criticality with peak performance, we observe that if there is a link between the two, it is rather loose. For $b<0$ the region of best performance is much broader than the critical region, indicated as hatched gray areas. In many cases, $b_{opt}$ does not lie within the critical region. For $b>0$, criticality and optimal performance seem more correlated, as optimal performance is usually obtained within the critical region. However, focusing on $K=4$, $b<0$ and the hardest memory task (Fig. \ref{fig:panel31}C), there is a striking difference between criticality and optimal performance: performance is almost zero in the critical region while it peaks in the region of re-entrance observed in the dynamics of the free running reservoirs, indicated in Fig. \ref{fig:panel31}C as an orange hatched area. Both regions show a variety of attractors, but only one corresponds to good performance. 

To conclude, in Fig.~\ref{fig:panel4}, we show a summary of the best performance in the memory (upper panel), and prediction (lower panel). In the plot, each dot represents the average over $20$ reservoirs obtained with the same connectivity parameters ($N$, $K$, $b_{opt}$), where $b_{opt}$ is the value that maximizes the average performance at the most difficult setting of each task ($\delta=-18$ and $\tau=28$), see Tab.~\ref{tab:1} and Tab.~\ref{tab:2}. As previously, we separated the case $b<0$ (left panel) and $b>0$ (right panel). We compare the performance for $K=1$ up to $16$. 

For all tasks, we note that the highest performances are consistently achieved with $K=3$ and $K=4$, irrespective of whether $b$ is positive or negative. However, the optimal value of $K$ exhibits some task dependency. In the memory task, for the more challenging task ($\delta=-18$), $K=4$ yields the best performance, despite $K=3$ occasionally outperforming less demanding tasks. This suggests that the optimal $K$ may depend on the complexity of the task at hand. The sign of $b$ has no discernible impact on the optimal $K$, however, it is observed that the performance for higher $K$ values is superior when $b>0$, in line with \citep{Calvet2023}. 

In the prediction task, again, the most challenging setup ($\tau=28$) shows $K=4$ as the optimal value, irrespective of the sign of $b$. In general, the reservoir-to-reservoir variance is very small for $b<0$. As previously observed, for higher $K$, we observe a significant reservoir variability, and this time, the performance is higher when $b<0$.

Taken together, these findings suggest that once an optimal value for $K$ is selected, the system's performance becomes mainly insensitive to the sign of the balance $b$, even though the optimal $K$ can be dependent on the task at hand.

\subsubsection{Discussion}

In line with \citep{Calvet2023}, for a positive balance, the critical region is reasonably aligned with the performing region, for all tested $K$. Yet our findings somewhat challenge the idea that the edge of chaos is always optimal for computation, as it does not necessarily overlap with the region of best performance. This is especially visible in the memory tasks and reservoirs with a negative balance. Indeed, for $K=4$, the re-entrant region provides the best reservoirs, while being very far from criticality.

By looking at dynamics, one might wonder if this re-entrant region of attractor diversity ($b<0$) does not belong to the critical region of the positive side, which, by shifting towards the left, overlaps on the negative sign. On the other hand, we observe a drastic dip in performance with both signs around $b=0$. This suggests that a breaking of symmetry is at play \cite{Coldenfeld2018}, acting as a crucial driver for performance while being surprisingly imperceptible in the dynamic.

Regarding reservoir design, we show that the optimal excitatory/inhibitory balance is intricately tied to the number of connections. For a high number of connections, a pronounced asymmetry is observed depending on whether there is a majority of inhibition or excitation. 

However, when $K=4$, the optimal $b$ value is almost identical and closely balanced between excitation and inhibition, regardless of whether $b$ is positive or negative. Consequently, the dynamics of reservoirs are nearly identical for both positive and negative $b$, resulting in similar performance outcomes. The task of choosing the optimal $b_{opt}$ becomes much simpler, as the asymmetry fades away.

\subsection{The interplay between reservoir size and connectivity degree} \label{sec:N}

This section studies the joint effect of the reservoir size $N$ (=100, 1000, 10000) and $K$, in relation to $b$. We show that $N$ has a comparable impact on the dynamics as $K$, but also impacts asymmetrically around $b$ the performance in tasks. 

\subsubsection{Impact of reservoir size and connectivity degree on dynamics} \label{sec:dynamic_K_and_N}

\begin{figure}
    \centering
    \includegraphics[width=1.\textwidth]{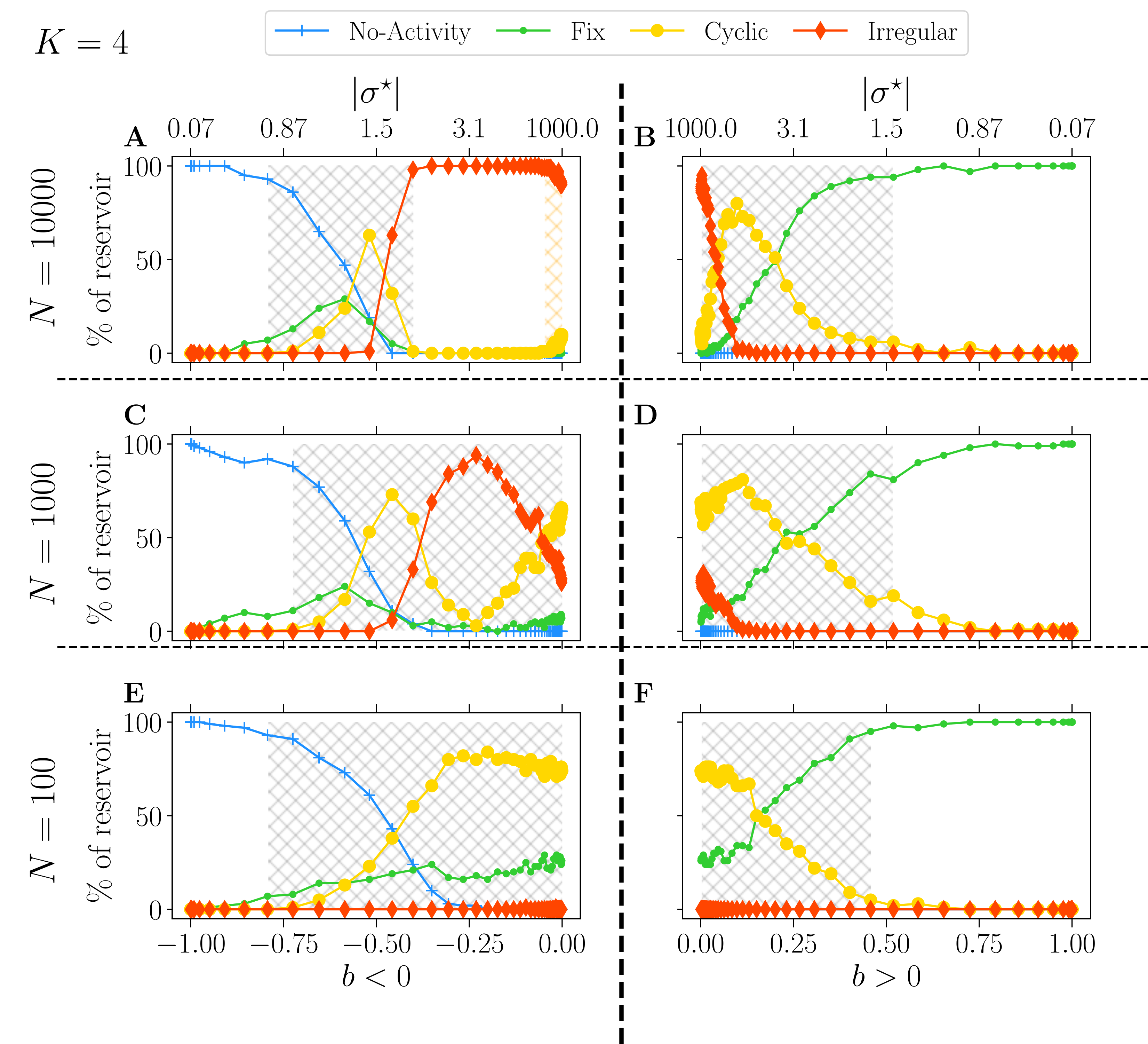}
    \caption{Attractor statistics of free-evolving RBN reservoirs for fixed $K=4$, with $N=10000$ (\textbf{A} and \textbf{B}), $N=1000$ (\textbf{D} and \textbf{D}), and $N=100$ (\textbf{E} and \textbf{F}). Statistics of attractors over 100 reservoirs run once (y-axis) versus $b$ (x-axis). The upper x-axis displays the corresponding $|\sigma^\star|$, both for $b<0$ (left panel) and $b>0$ (right panel). Each activity signal is classified into one of the six categories of attractors: extinguished, fixed, cyclic, and irregular, defined in methodology Sec.~\ref{met:experiments}. The light-gray hatched areas represent the critical regions defined; the threshold is chosen to $0.0001$. In \textbf{A}, the orange hatched area represents the region of re-entrance of criticality with non-zero BiEntropy variance.}
    \label{fig:panel5}
\end{figure}

In Fig.~\ref{fig:panel5}, we set $K=4$ and present the attractor statistics over $b$ for three different values of $N$: $N=10000$ (upper panel), $N=1000$ (middle panel), and $N=100$ (lower panel). We analyze these values in two cases, $b<0$ (left panel) and $b>0$ (right panel).

From our observations, it is evident that reducing $N$ leads to a decrease in the complexity of the attractors, as indicated by the reduction of irregular attractors. In the case of $b<0$ and as $N$ decreases, the re-entrant region (orange hashed area) observed with $N=10000$ (Fig.~\ref{fig:panel5}.\textbf{A}) merges with the critical one (gray hashed area) for $N=1000$ (Fig.~\ref{fig:panel5}.\textbf{C}), resulting in a spike of irregular attractors and eventually leaving room for predominantly cyclic ones as $N=100$ (Fig.~\ref{fig:panel5}.\textbf{C}).

Contrarily, for $b>0$ and $N=1000$, this spike or irregular attractor is missing, and the critical phase is largely dominated by cyclic attractors, with only a few fixed and irregular ones. Interestingly, when $N=100$, both signs yield very similar results, with no irregular attractors at all. This observation underscores the impact of $N$ on the nature and complexity of the attractors.

Lastly, when discussing Fig.~\ref{fig:panel2}, we briefly mentioned the continuity in attractor statistics as going from a negative to a positive balance. This fact is even more salient in Fig.~\ref{fig:panel5}. Statistics of attractors closely match on both sides, reinforcing the picture that the critical region can span both signs, at least from the dynamic lens. 

\subsubsection{Impact of reservoir size and connectivity degree on performance} \label{sec:performance_K_and_N}

\begin{figure}
    \centering
    \includegraphics[width=1\textwidth]{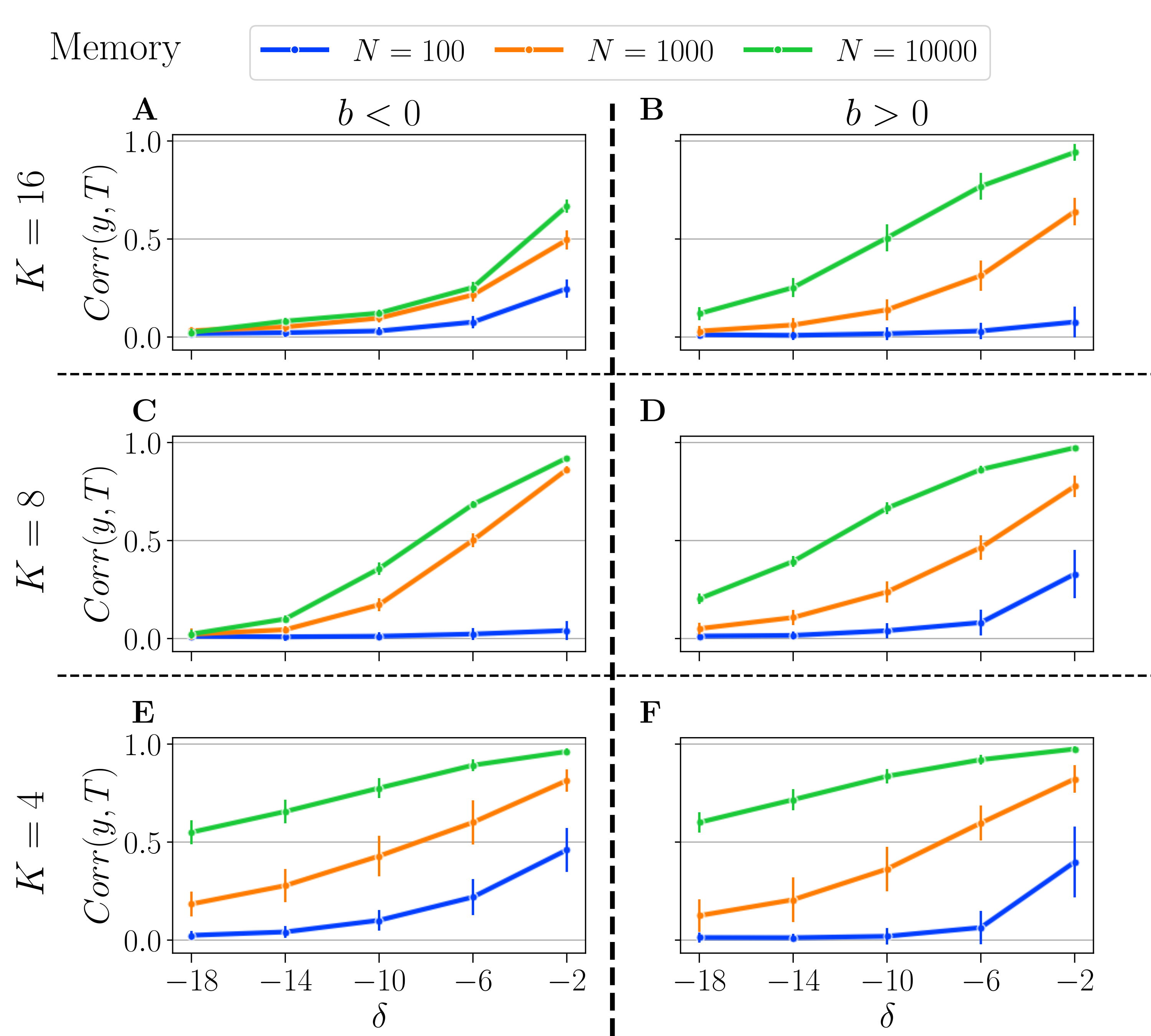}
    \caption{Summary of performance in the memory tasks, for various connectivity degrees $K$, and size of the reservoirs $N$: for $N=10000$ (green curves), $N=1000$ (orange curves), and $N=100$ (blue curves). $K=16$ (\textbf{A} and \textbf{B}), $K=8$ (\textbf{C} and \textbf{D}),  $K=4$ (\textbf{E} and \textbf{F}). $b<0$ (left column), and $b>0$ (right column). Solid lines represent the average over all reservoirs generated with the same reservoir ($N$, $K$, $b_{opt}$), and the error bar represents one standard deviation. As explained in Sec.~\ref{sec:perf_K_and_b}, $b_{opt}$ is obtained by selecting the balance that gives the best average performance at the most difficult setting in each respective task.}
    \label{fig:panel6}
\end{figure}

\begin{figure}
    \centering
    \includegraphics[width=1\textwidth]{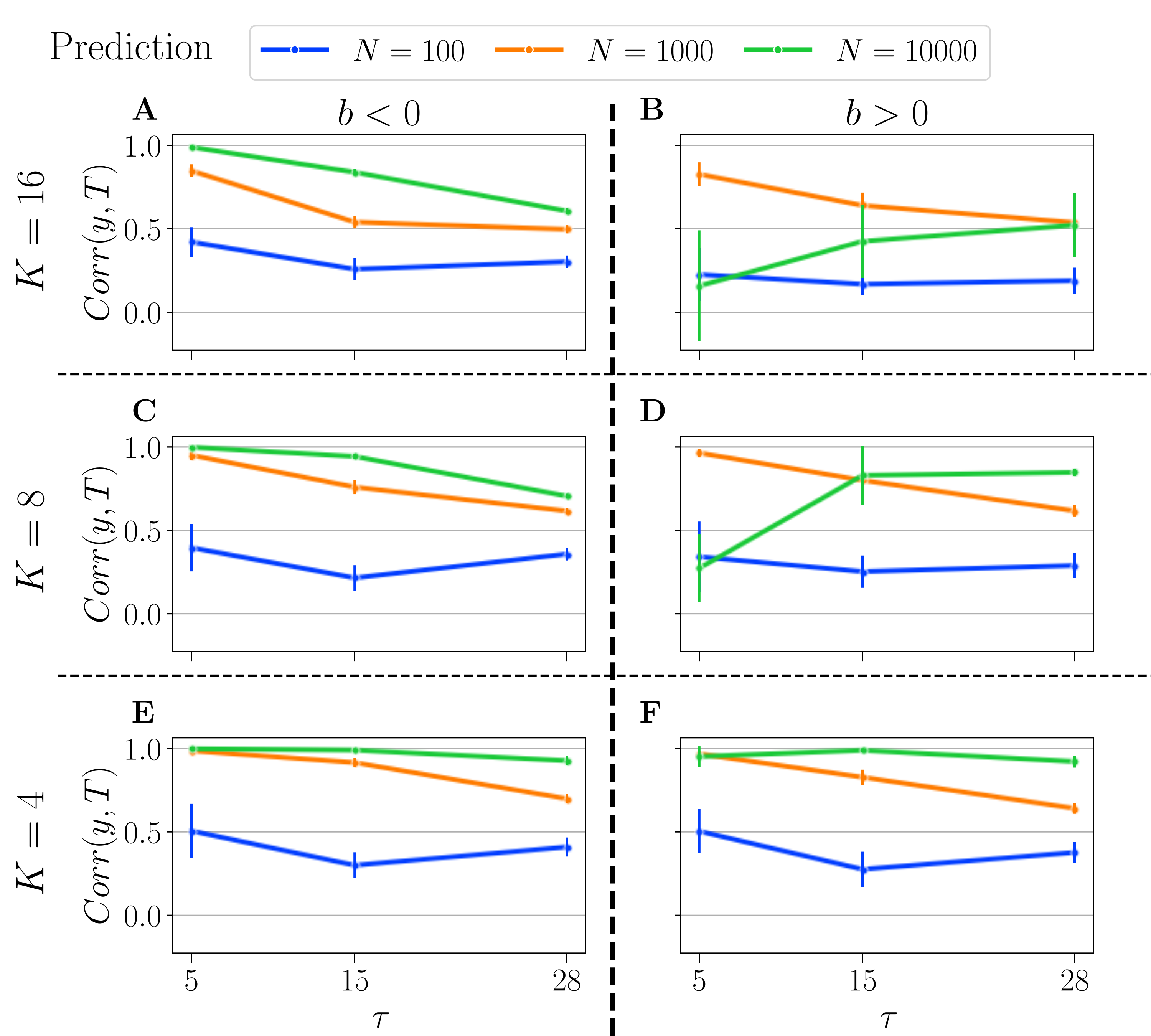}
    \caption{Summary of performance in the prediction tasks, for various connectivity degrees $K$, and size of the reservoirs $N$: for $N=10000$ (green curves), $N=1000$ (orange curves), and $N=100$ (blue curves). $K=16$ (\textbf{A} and \textbf{B}), $K=8$ (\textbf{C} and \textbf{D}),  $K=4$ (\textbf{E} and \textbf{F}). $b<0$ (left column), and $b>0$ (right column). For more information on the plots, see the caption of Fig.~\ref{fig:panel6}.}
    \label{fig:panel7}
\end{figure}

Results for the memory task and prediction are respectively displayed in Fig~\ref{fig:panel6} and Fig.~\ref{fig:panel7}. We tested the performance for $K=4$ (upper panel), $K=8$ (middle) and $K=16$ (bottom). Reservoirs with $b<0$ are displayed in the left panel and $b>0$ in the right panel. We compare the performance for three distinct values of $N$: $N=10000$ (green curves), $N=1000$ (orange curves), and $N=100$ (blue curves). As in the previous Sec.~\ref{sec:perf_K_and_b}, performance is shown for $b_{opt}$, established for the most difficult setting in each task ($\delta=-18$ and $\tau=28$).

In the memory task, as expected, reducing the number of neurons diminishes the reservoirs's memory capacity, and the more difficult the task, the lower the performance. In addition, decreasing the reservoir size generally increases the reservoir-to-reservoir variance, as indicated by the larger error bars, even though this is not always the case, especially when performance is already low. 

The number of neurons exerts a greater influence when $K$ is lower. Indeed, for $K=4$, we observe a significant disparity between all three $N$ values across all difficulty levels ($\delta$). Surprisingly, for higher $K$, and especially when $b<0$, performances for $N=1000$ and $N=10000$ are relatively comparable, and increasing the reservoir size is not improving performance, especially for tasks requiring longer memory. As previously noted, when $K=4$, performance is similar regardless of whether $b$ is positive or negative, a finding that is now corroborated across all tested $N$ values. 

It appears that $K$ has minimal influence when $N=100$, as reservoirs perform similarly regardless of $K$. The same holds true for both positive and negative $b$, which their identical dynamics profiles might explain. This suggests that, for low neuron counts, the system's dynamics and performance are more strongly influenced by the balance parameter $b$ than by the number of connections $K$. 

In the prediction task, we observe some surprising trends. Notably, having a higher $N$ is not always advantageous, as the optimal $N$ appears to depend on both the task and the control parameter.

Firstly, for $b<0$, the performance profile is similar to that in the memory task: higher $N$ yields better performance, and performance decreases with increasing task difficulty ($\tau$). However, the performances of $N=10000$ and $N=1000$ are closer to each other and significantly higher than that of $N=100$, which again remains unaffected by $K$.

Secondly, for $b>0$, the value of $K$ strongly influences the relationship between performance and reservoir size. With $K=4$, the performance profile is similar to that for $b<0$: performance decreases monotonically with $\tau$ and $N$. However, for $K=8$ and especially for $K=16$, we observe some unexpected results. Smaller reservoirs ($N=1000$) can outperform larger ones ($N=10000$) in some tasks. This phenomenon is even more pronounced for higher $K$, as the orange line (representing $N=1000$) consistently outperforms the green line (representing $N=10000$) across all tested tasks.

%% file: conclusion.tex
\vspace{10pt}
\section{Discussion}\label{Sec:Conclusion}

Our study reveals that the edge of chaos, or the critical region, does not consistently align with the peak performance region \citep{Gallicchio2020}, and this alignment is contingent upon the sign of excitatory-inhibitory balance $b$. For $b>0$, as previously observed \citep{Calvet2023}, the critical region coincides with the highest performance. However, for $b<0$, the region of optimal performance does not coincide with the critical region when the connectivity degree $K$ is optimally selected. Instead, supplanting the disordered phase, a re-entrance of the critical region is observed, indicated by an increased attractor diversity, which surprisingly aligns with the best-performing region. This insight suggests that the attractor dynamics can be utilized to identify the region of interest for the design or reservoirs, and this also holds for $b>0$ and its identified critical region \citep{Calvet2023}. 

In terms of the interplay between $b$ and the connectivity degree $K$, our research shows that a carefully selected $K$ ($K=4$) renders the sign of $b$ irrelevant, as the optimal $b$ becomes $\pm \epsilon$ with $\epsilon$ very small. This suggests that the optimal balance is near, but not at, perfect symmetry, even though $b\rightarrow0$ results in zero performance. In statistical physics, it is well known that symmetry breaking induces critical phase transitions \citep{Coldenfeld2018}, and our findings suggest that symmetry breaking in the balance of excitatory-inhibitory synapses is crucial for achieving optimal performance. Refining initial literature \citep{Bertschinger2004c, Snyder2013a, Burkow2016, Echlin2018}, the highest-performing region is characterized by a preponderance of irregular attractors within the disordered region.

To understand this, one can consider what happens when $\sigma^*$ tends to infinity. This can be achieved in two ways: first, when the standard deviation of the weight $\sigma$ is fixed while the mean weights $\mu\rightarrow0$, and second, when $\mu$ is fixed while $\sigma\rightarrow \infty$. The first case has been covered in other works \citep{Bertschinger2004c, Busing2010} and shows the importance of tuning the scaling of the input weights with the recurrent weight statistics \citep{Burkow2016}. In the present work, however, the second option is considered, as the mean weights is fixed, and $\sigma$ increases to higher values. As such, $b$ approaches zero, which results in a symmetry between excitation and inhibition but with increasingly higher synaptic weights (in absolute value). Consequently, each neuron receives equal excitatory and inhibitory recurrent inputs, and since the input weights are kept constant, the external input becomes insignificant. Finally, since neurons have a zero threshold, they have a 50\% probability of spiking, leading to a random spike train. Therefore, it is not surprising to observe a performance dip as $\sigma \rightarrow\infty$ ($b \rightarrow 0$) since the reservoir activity becomes independent of the input. However, what requires further investigation is the unexpected drastic performance increase when this symmetry is slightly broken as $b = \pm \epsilon \sim 0.03$ (roughly corresponding to a 6\% difference between excitatory and inhibitory synapses). 

These findings highlight the critical role of $K$ in determining other control parameters. Firstly, the optimal number of connections ($K=4$) eliminates the performance asymmetry, significantly simplifying the parameter $b$ selection. Secondly, consistent with previous studies, $N$ generally enhances performance, but this is only true for optimal $K=4$ values, particularly in the prediction task, where smaller reservoirs occasionally outperform larger ones. Additionally, the performance gain obtained by $K$ is significant only when the reservoir size is sufficiently large. For instance, with reservoirs of size $N=100$, $K$ had close to no effect on the best performance. However, optimally choosing $K$ becomes key to obtaining a gain in performance when increasing the size.

Our work reveals a complex interplay between the topology and weights parameters, but assuming a reservoir of sufficient size ($N\ge 1000$), $K$ acts as a pivotal control parameter by greatly simplifying the way parameters interact with each other. When $K$ is optimal, then $N$ must be maximized, and $b$ can be chosen very close to zero but finite, and of any sign.

\section{Future work}\label{Sec:future}

Understanding the relationship between dynamics and performance is crucial for simplifying reservoir design \citep{Bertschinger2004c, Krauss2019, Krauss2019a, Metzner2022, Calvet2023}. Our study advances this understanding but reveals a more intricate relationship than anticipated. Specifically, performance was found to be highly sensitive to symmetry breaking in the excitation-inhibition balance, while the metrics used to probe the dynamics were completely unaware of the symmetry. Future research could investigate the relationship between the optimal balance and other dynamic-probing metrics, including spatial and temporal correlation \citep{Metzner2022}, and possibly topology \citep{Kinoshita2009, Masulli2016}. For instance, it could be hypothesized that the longest neural pathways in the random graph become available for information transmission only at $b=\pm \epsilon$, which could explain why optimal performance necessitates a breaking of symmetry in the balance.

%% file: supplementary.tex
\section*{Supplementary Material}
\beginsupplement

\section{Training} \label{suppm:training}

Both tasks follow the same protocol for training the readout weights:
\begin{itemize}
    \item Reservoirs receive the input $u$ for $D=2000$ time steps.
    \item We discard the first $500$ time steps.
    \item The training is then performed on the subsequent $1500$ time steps. We concatenate the reservoir outputs in time, and use the optimization procedure defined in \ref{met:model}.
\end{itemize}
\noindent Each experiment consists in $40$ values of $\sigma^{\star}$ per sign of $b$, and is performed for three values of $K=\{4, 8, 16\}$ and three values of $N=\{100, 1000, 10000\}$. For each value ($\sigma^{\star}$, $K$, $N$), $20$ reservoirs are randomly generated, and each network is run $5$ times with different randomly tossed inputs (i.e., $14,400$ simulations). Each training is performed for $4000$ epochs (with a total of $57,600,000$ training epochs). 

\subsection{The density is not a control parameter} \label{suppm:density}

Historically, the litterature on RBN studied the connectivity degree $K$, along with the number of neurons in the reservoir $N$. It has been shown that RBN reservoirs possess interesting computational properties for very sparse matrices, with at most $K=25$ connections per neuron, according to \citep{Busing2010}. On the other hand, studies in ESN have explored the impact of another parameter, the density $d$ \citep{Hajnal2006, Krauss2019a, Metzner2022}, which captures how many zeros there are in the adjacency matrix, ranging from zero for no connections, to one for fully connected reservoirs. 

In the context of RBN reservoirs, the density is equal to $d=K/N$. It is worth noting that for the values of $K$ and $N$ used in this study, the corresponding values of $d$ are ridiculously small, ranging from $0,0004$ to $0.16$. This stresses how much the RBN behaves differently since phase transitions occur at a fraction of what is observed in an ESN. It also suggests that $K$ is a more natural choice for the control parameter of the RBN.

Nonetheless, this is not sufficient to rule out $d$ as a good candidate for control parameters. Recalling that it is in the same fashion that $\mu$ and $\sigma$ were shown to be dependent parameters and $\sigma^\star=\mu/\sigma$ was built in \citep{Calvet2023}. If $d$ is a control parameter, however, then all combinations of $K$ and $N$ leading to the same values of $d$ should provide the exact same dynamics and performance in tasks, as is the case with $\sigma^\star$. To find if $d$ is indeed a control parameter, we study the dynamics and performance of three couples of $K$ and $N$ giving the same value $d$:

$$d=K/N=16/10000=8/5000=4/2500=0.0016$$

\begin{figure}
    \centering
    \includegraphics[width=1\textwidth]{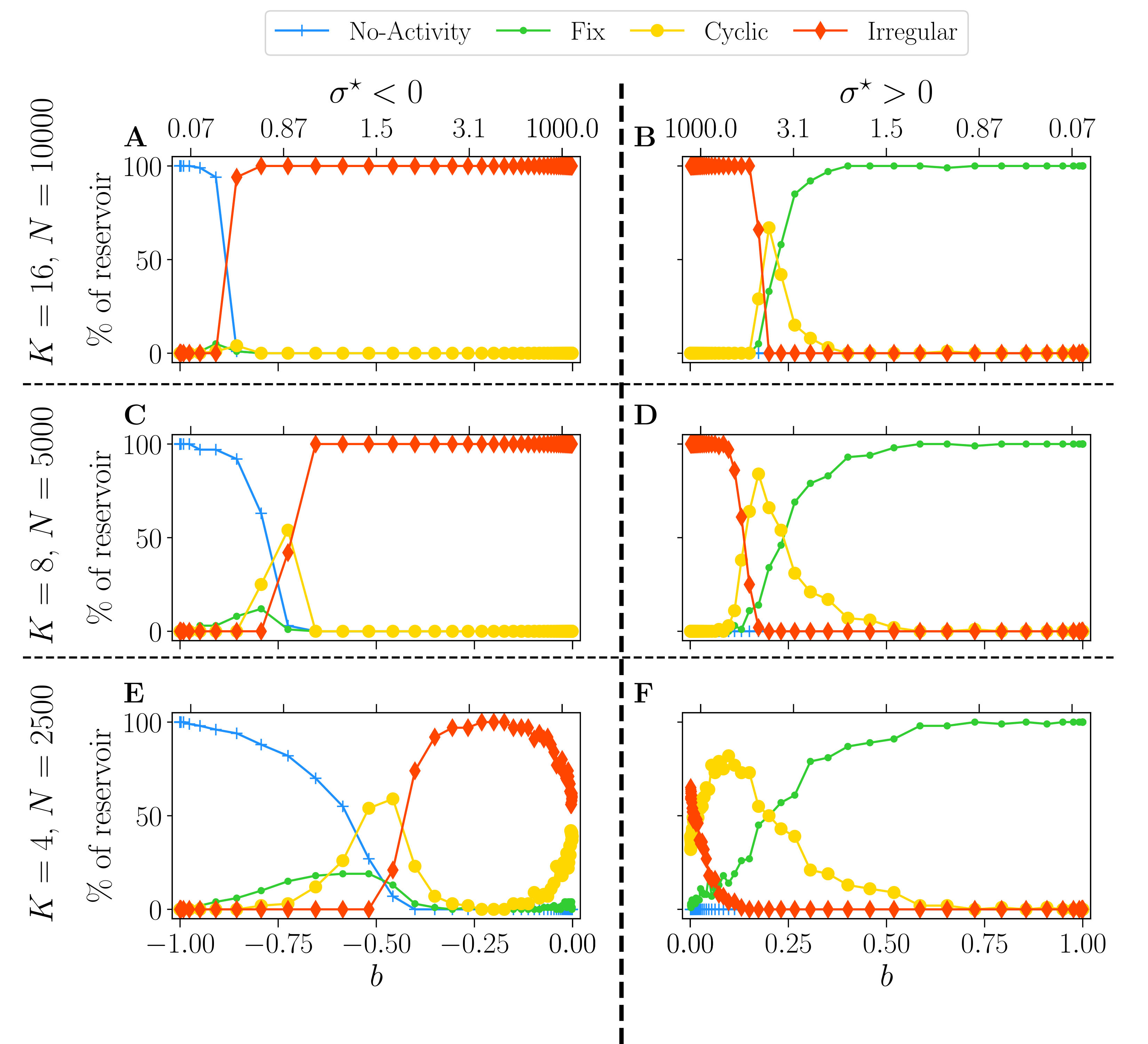}
    \caption{Attractor statistics of free-evolving RBN reservoirs, controlled by $K$ and $N$ with a constant ratio (rows), versus the balance $b$ (x-axis). The lower x-axis represents the corresponding $|\sigma^\star|$, for $b<0$ (\textbf{A}, \textbf{C}, \textbf{E}), and $b>0$ (\textbf{B}, \textbf{D}, \textbf{F}). Each steady activity signal is classified into one of the four categories of attractors: no-activity (\textcolor{dead}{$+$}), fix (\textcolor{fix}{$\sbullet$}), cyclic (\textcolor{cyclic}{$\bullet$}), irregular (\textcolor{chaos}{$\blacklozenge$}). The statistics of attractors are computed over 100 reservoirs run once (y-axis). Results are shown for $K=16$, $N=10000$ (\textbf{A} and \textbf{B}), $K=8$, $N=5000$ (\textbf{C} and \textbf{D}), and $K=4$, $N=2500$ (\textbf{E} and \textbf{F}).}
    \label{fig:s1}
\end{figure}

In Fig.~\ref{fig:s1}, we show the attractor statistics in the three conditions, as a function of $b$. The first thing we notice is that all three tested conditions give very different attractor profiles. The width and the location of the critical regions are not the same, and the proportion of attractors also greatly varies, with lower $K$ and $N$ giving much more varied dynamics and wider critical regions. It is intriguing that in Fig.~\ref{fig:panel2} and Fig.~\ref{fig:panel5}, diminishing $K$ and $N$ results in a somewhat comparable increase in the dynamics variability as well, but if $K$ and $N$ seem related, it is not by their ratio as proposed in the density parameter. 

\begin{figure}
    \centering
    \includegraphics[width=1\textwidth]{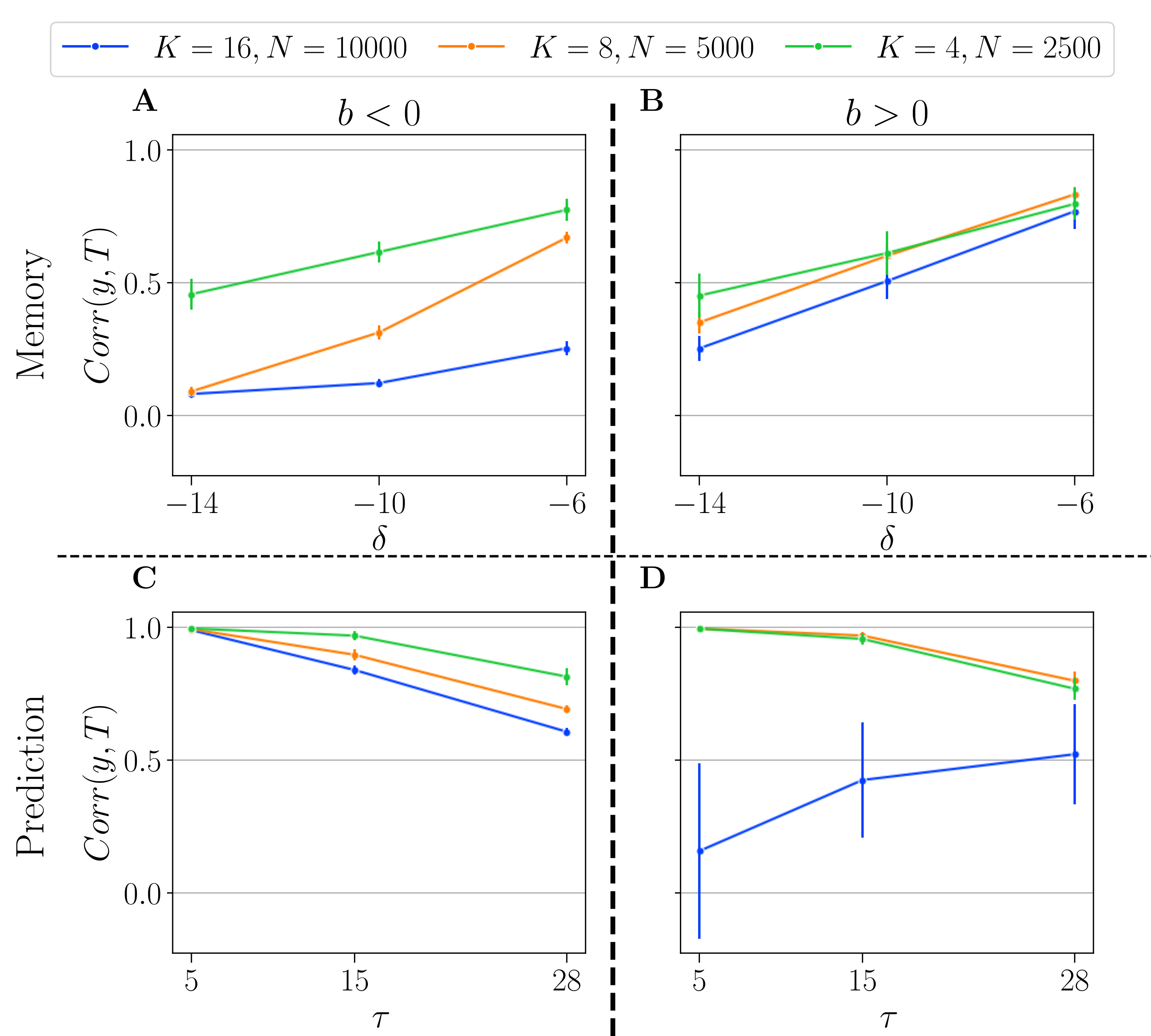}
    \caption{Summary of performance in the memory (\textbf{A} and \textbf{B}) and the prediction (\textbf{C} and \textbf{D}) tasks, for three couples of $K$ and $N$ resulting in the same density: $K=4, N=2500$; $K=8, N=5000$; and $K=16, N=10000$. Results are shown for both $b<0$ (left panel), and $b>0$ (right panel). For each value of ($K$, $N$), we selected the $b_{opt}$ giving the highest average performance, in the most difficult task ($\delta=-14$ for memory, and $\tau=28$ for prediction). We plot the performance (higher is better) of reservoirs $Corr(y, T)$ (y-axis), plotted as a function of their respective task parameters (x-axis): $\delta$ (memory) and $\tau$ (prediction). The solid line represents the average over $20$ reservoirs (generated with the same $b_{opt}$ and ($K$, $N$) values. The error bars represent one standard deviation.}
    \label{fig:s2}
\end{figure}

In addition, we show in Fig.~\ref{fig:s2} the performance in memory and prediction of $20$ reservoirs selected with the same $b_{opt}$ values, for each couple ($N$, $K$) (for more information on how these plots are obtained see Sec.~\ref{sec:perf_K_and_b}). Following the dynamics analysis, even though each couple ($N$, $K$) possesses the same density, the performance varies. 

When considering $b<0$, smaller $K$ consistently provides higher performance in all considered tasks. This is interesting because it is in spite of the fact that reservoirs are smaller. On the other hand, for $b>0$, we have a different behaviour, first in the memory task, performances are close even though ($K=4$, $N=2500$) performs slightly better in the most difficult ($\tau=-14$) set-up. Second, in the prediction task, ($K=4$, $N=2500$) and ($K=8$, $N=5000$) give very similar performance, while the latter have reservoirs of twice the size. Surprisingly, even though ($K=16$, $N=10000$) has the biggest network, its performance is way below the two others. 

As a consequence, for the same given density value obtained with various combinations of $K$ and $N$, we obtain distinct dynamics statistics and performance. We conclude that the density is not a control parameter for RBN reservoirs, and one must study the effect of $K$ and $N$ separately, as two independent control parameters. 

Regarding the objective of the paper, these results suggest that $K$ has a stronger influence on the performance than $N$, as increasing the size of the network does not necessarily guarantee higher performance, but reducing $K$ consistently improves them. 

\section{Optimal balances for performance} \label{suppm:b_opt}

\begin{table}[h]
    \centering
    \begin{tabular}{l|lllllll}
        \hline
    $K$   & 1  & 2  & 3  & 4  & 5 & 8  & 16  \\ \midrule
    $b>0$ & 0.024  & 0.024  & 0.052 & \cellcolor{green!25} 0.029  & 0.077  & 0.149 & 0.179  \\ 
    $b<0$ & -0.099 & -0.069 & -0.057 & \cellcolor{green!25} -0.029 & -0.049 & -0.52 & -0.99 \\ \bottomrule
    \hline
    \end{tabular}
    \caption{Table of the balance $b_{opt}$ giving the best average performance at white-noise memory for $\delta=-18$ ($N=10000$). The green highlight represents the optimal $K$ value. Lower values of $K$ typically give the lowest $b_{opt}$ values, while increasing $K$ tends to increase $b_{opt}$ in absolute value.}
    \label{tab:1}
\end{table}

\begin{table}[h]
    \centering
    \begin{tabular}{l|lllllll}
        \hline
    $K$   & 1  & 2  & 3  & 4  & 5 & 8  & 16  \\ \midrule
    $b>0$ & 0.95  & 0.19  & 0.068 & \cellcolor{green!25} 0.024  & 0.098 & 0.15  & 0.19  \\ 
    $b<0$ & -0.24 & -0.20 & -0.12 & \cellcolor{green!25} -0.047 & -0.42 & -0.82 & -0.95 \\ \bottomrule
    \hline
    \end{tabular}
    \caption{Table of the balance $b_{opt}$ giving the best average performance at Mackey-Glass prediction for $\tau=28$ ($N=10000$). The green highlight represents the optimal $K$ value, which typically gives the lowest $b_{opt}$ values, while increasing $K$ tends to increase $b_{opt}$ in absolute value.}
    \label{tab:2}
\end{table}